\documentstyle[12pt]{article}
\newcommand{\DATE}  {\today}

\newcommand{\PPrtNo}
{
MSU-HEP-61222 \\ CTEQ-622 
}
\newcommand{\TITLE}
{
Charm Production and Parton Distributions
}
\newcommand{\THANKS}
{
This work is partially supported by the National Science Foundation.
}
\newcommand{\AUTHORS}
{
H.L. Lai and W.K. Tung
}
\newcommand{\INST}
{
Michigan State University, E. Lansing, MI 48824, USA
}
\newcommand{\ABSTRACT}
{ Recent accurate data on $F_2(x,Q)$ and on $F_2^c(x,Q)$ from HERA at small-$x$
require a more precise treatment of charm production in the global QCD analysis
of parton distributions.  We improve on existing global analyses by
implementing the leptoproduction formalism of Aivazis \etal which represents a
natural generalization of the conventional zero mass QCD parton framework to
include heavy quark mass effects.  We also perform analyses based on the
fixed-flavor-number scheme, which is widely used in the literature, and
demonstrate their uses and limitations. We discuss the implications of the
improved treatment of heavy quark mass effect in practical applications of PQCD
and compare our results with recent related works.  }

\newcommand{\ie}{{\it i.e.,\ }}
\newcommand{\eg}{{\it e.g.,\ }}

\newcommand{\etal} {{\it et al.\ }}

\def\stackunder#1#2{\mathrel{\mathop{#2}\limits_{#1}}}
\input epsf
\newcommand{\epsfile}[1]{\epsfbox{#1}}
\def\figdiagram
{
\begin{figure}[htb]
\epsfxsize=5.2in
\centerline{\epsfile{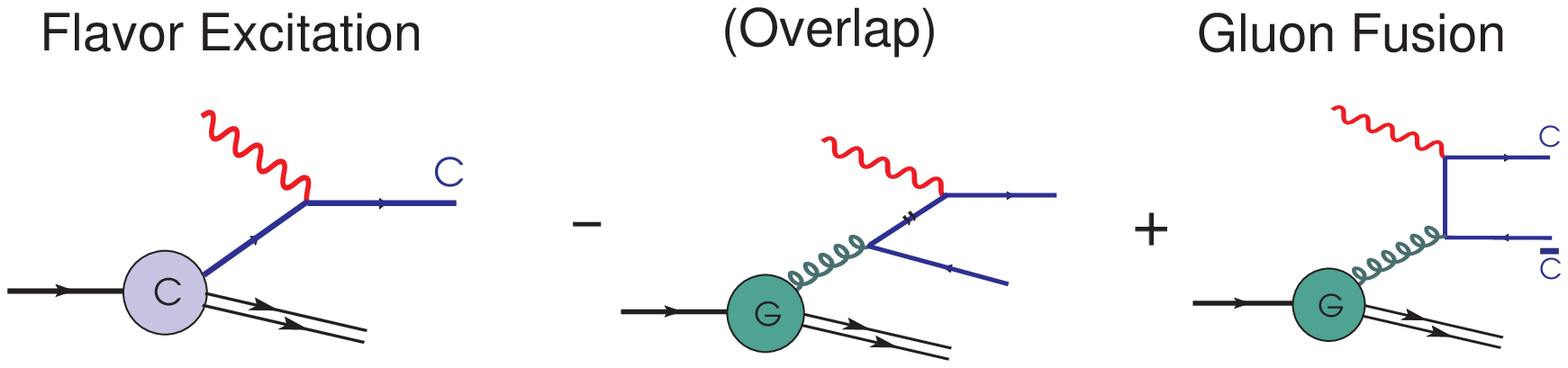}}
  \caption{Charm production in NC deep inelastic scattering in PQCD: flavor
excitation and gluon fusion production mechanisms and their overlap 
which has to be subtracted. The mark ${}_{''}$ on the internal quark line
indicates it is on-mass-shell and collinear to the gluon.}
  \label{fig:diagram}
\end{figure}
}
\def\fighera
{
\begin{figure}[htb]
\epsfysize=3.2in
\centerline{\epsfile{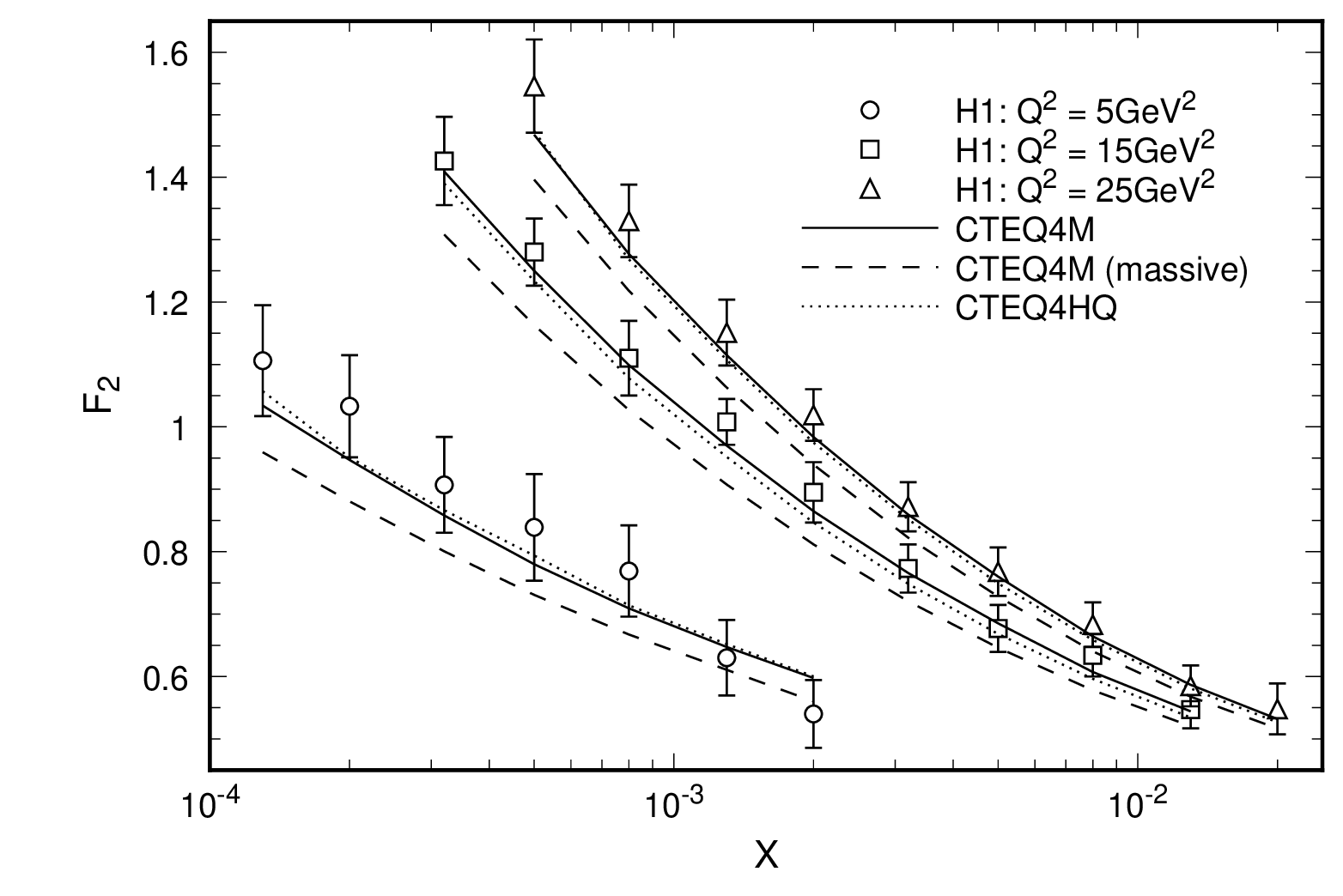}}
  \caption{Comparison of H1 data in the small-$x$ region with calculations using 
CTEQ4M parton distributions in the (original) massless QCD scheme (solid line) and with massive hard cross-section formulas of Sect. 2 (dashed line). 
Also shown is the result of the new fit CTEQ4HQ.}
  \label{fig:hera}
\end{figure}
}

\def\figpercent
{
\begin{figure}[htb]
\epsfysize=3.2in
\centerline{\epsfile{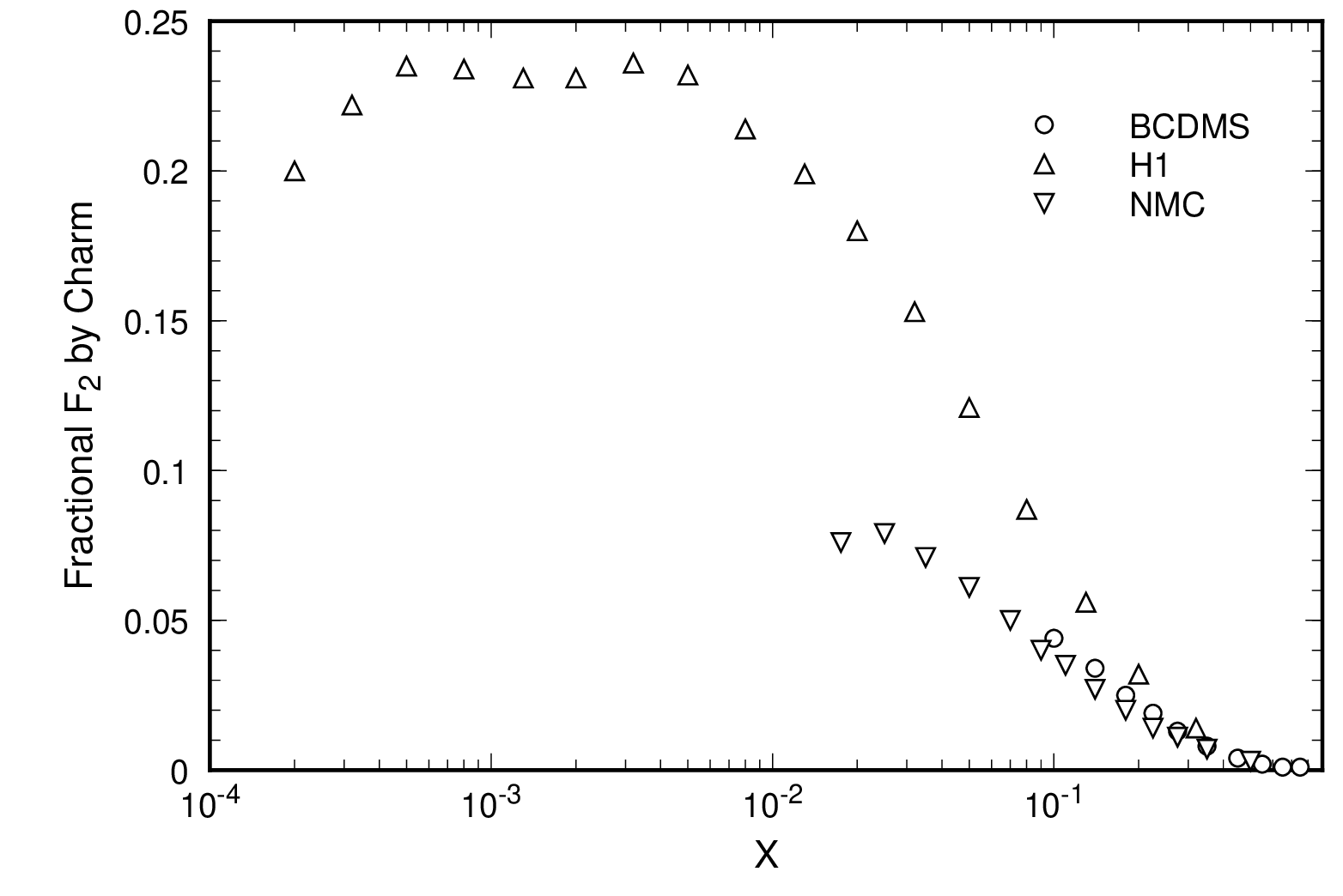}}
  \caption{Fractional contribution on $F_2$ from the charm quark for different experimental ($x$,Q) range. The data shown are calculated using CTEQ4M with average on the Q bins of that particular experiment at fixed $x$.}
  \label{fig:percent}
\end{figure}
}
\def\figpdf
{
\begin{figure}[hbt]
\epsfysize=3.2in
\centerline{\epsfile{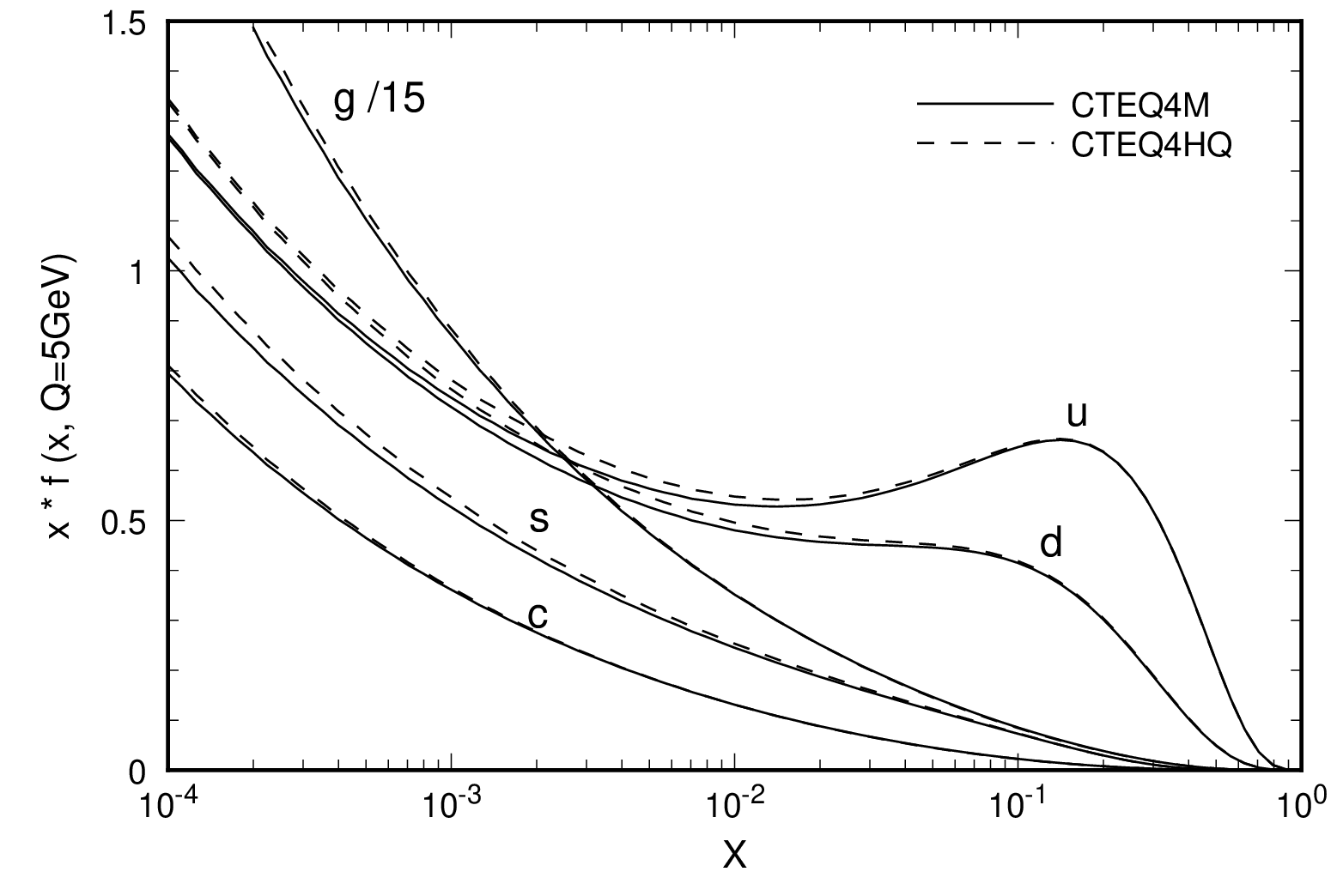}}
  \caption{Comparison of CTEQ4M and CTEQ4HQ parton distributions.}
  \label{fig:pdf}
\end{figure}
}
\def\tblchi
{
\begin{table}[htb]
\begin{center}
\begin{tabular}{|c|c||c|c|c|c|c|}
\hline
 Expt.     &  \#pts &  CTEQ4M   &  CTEQ4HQ   &  CTEQ4F4  &  CTEQ4F3        \\
\hline\hline
$FT-NC$    &   691  & 725 & 716 & 716 & 721 \\ \hline
$CCFR$     &   126  & 130 & 125 & 120 & 135 \\ \hline
$HERA$     &   351  & 362 & 350 & 409 & 410 \\ \hline
$CDF A_W$  &     9  &   4 &   5 &   4 &  11 \\ \hline
$NA51$     &     1  & 0.6 & 0.7 & 0.8 & 0.7 \\ \hline
$E605$     &   119  &  98 &  96 &  99 & 101 \\ \hline
\hline
 Total     &  1297  &   1320    &   1293    &   1349    &   1380    \\ \hline
\end{tabular}
\end{center}
  \caption{Total $\chi^2$ values and their distribution among the DIS and DY 
experiments: The first column $FT-NC$ stands for fixed target neutral current experiments which includes BCDMS, NMC and E665.}
  \label{tbl:chi}
\end{table}
}
\begin{document}

\begin{tabular}{l}
\DATE
\end{tabular}
\hfill
\begin{tabular}{l}
\PPrtNo
\end{tabular}

\vspace{1cm}

\begin{center}
\renewcommand{\thefootnote}{\fnsymbol{footnote}}
{
\LARGE \TITLE  \footnote[2]{\THANKS}
}
\renewcommand{\thefootnote}{\arabic{footnote}}

\vspace{1cm}
{\large  \AUTHORS}

\vspace{0.5cm}
\INST   \\

\vspace{0.5cm}
Abstract 

\end{center}

\ABSTRACT                 

\section{Introduction}

The QCD parton picture, based on perturbative Quantum Chromodynamics (PQCD),
has furnished a remarkably successful framework to interpret a wide range of
high energy lepton-lepton, lepton-hadron, and hadron-hadron processes.
Through global analysis of these processes, detailed information on the
parton structure of hadrons, especially the nucleon, has been obtained. \
Existing global analyses have mostly been performed within the traditional
zero-mass parton picture since, up to now, ``heavy quark'' partons (charm,
bottom and top) have played a relatively minor role in physically measured
quantities used in the analyses. With the advent of precise data on
inclusive $F_2$ \cite{hera} and the direct measurement of the charm
component $F_2^c$ \cite{H1c} from HERA, this is no longer the case. The
latter comprises about 25\% of the inclusive structure function at small-$x$%
. It is now necessary to sharpen the formulation of the theory for heavy
flavor production used in these global analyses.

The conventional QCD parton model is formulated in the zero-mass parton
limit. Since the early 1980's, when the parton picture became the essential
tool in calculating all high energy processes in the Standard Model and in
the search for ``New Physics'', it has been generalized to the form of a
master (factorization) formula 
\begin{equation}
\sigma _{N\rightarrow X}(s,Q)=\stackunder{a}{\sum }f_N^a(x,\mu )\otimes \hat{%
\sigma}_{a\rightarrow X}(\hat{s},Q,\mu )  \label{factorlN}
\end{equation}
where $N$ denotes one or two hadrons in the initial state; $X,$ a set of
inclusive or semi-inclusive final states consisting of ordinary or new
particles; ``$a$'', the parton label; $f_N^a(x,\mu ),$ the parton
distribution at the factorization scale $\mu $; $\hat{\sigma}_{a\rightarrow
X},$ the perturbatively calculable partonic cross-section; and the parton
label ``$a$'' is to be summed over all possible {\em active parton species}.
``Active'' partons, according to this widely accepted credo, include all
quanta which can participate effectively in the dynamics at the relevant
energy scale\cite{marciano,coltun}, here denoted generically by $Q$ (\eg $Q$
in deep inelastic scattering, $M_W$ in $W$ production, $p_t$ in direct
photon or jet production, ... ). Thus, for heavy particle production ($%
W,\;Z,\;Higgs,\;SUSY\;particles...$), $a=\left\{
g,\,u,\,d,\,s,\,c,\,b,(t)\right\} .$ This general picture has been adopted
by most global analysis work\cite{ehlq,mrs,cteq}, resulting in several
generations of parton distributions which include all the parton flavors.
The charm, bottom (and sometimes top) quark distributions all contribute to
high energy processes above energy scales higher than the respective quark
masses.

Viewed from this perspective, the QCD theory for heavy quark flavor
production poses a special challenge. If a heavy quark, say charm $c,$%
\footnote{%
For clarity, and for the specific applications of this paper, we shall focus
on charm in most of our discussions. The same considerations apply to bottom
quark production with the substitution $c\rightarrow b$ everywhere.} is
produced as part of the final state $X,$ (i) should one treat this process
literally like the production of other heavy particles ($W,\;Z,\;Higgs,...$%
), \ie differentiate $c$ from the other partons, and exclude it from the
initial state (by restricting the sum over ``$a$'' Eq.~\ref{factorlN} to
only the light quarks)\cite{GHR,GRV} ; or, (ii) based on the very physical
ideas behind the factorization formula, should one still count $c$ among the
initial state partons because (unlike the electroweak heavy particles) it
certainly is an active participant in strong interaction dynamics, including
its own production, provided the energy scale is high enough \cite
{coltun,acot}? In principle, the two alternatives can be regarded as two
different but equivalent {\em schemes} for organizing the perturbation
series in PQCD. In practice, since the perturbation series is truncated
after one or two terms, the effectiveness (\ie accuracy) of the two
approaches can be quite different in different kinematic regions.

Much of the recent specific literature on heavy quark production is based on
the first approach ((i) above)\cite{nde,smithetal}, extended to
next-to-leading order (NLO), including in the initial states only the light
partons---$\left\{ g,\,u,\,d,\,s\right\} $ for charm production. These are
referred to as {\em fixed-flavor-number} (FFN) calculations---3-flavor ($%
n_f=3$) for charm. This scheme is conceptually simple. However, when the
ratio of energy scales $\frac Q{m_c}$ becomes large, \ie when the charm
quark becomes relatively light compared to the prevalent energy scale, its
reliability comes into question because of the presence of large $\ln \frac
Q{m_c}$ factors---both in terms kept and terms left out. For measurements at
HERA (and for both charm and bottom production at the Tevatron), this
consideration becomes increasingly relevant. In fact, tell-tale signs of the
inadequacy of these calculations have been known for some time: (i) the
calculated cross-sections have unacceptably large dependence on the
renormalization and factorization scale $\mu ;$ and (ii) the theoretical
cross-sections, in most cases, fall well below the experimental values for
all reasonable choices of $\mu .$

On the other hand, as mentioned above, most work on global analysis of
inclusive structure functions (which contain heavy quark final state
contributions) use the second approach -- but with a simplification: once a
massive quark is turned on above its threshold, it is treated as massless,
on the same footing as the other light flavors. This is an approximation
which is strictly correct only in the asymptotic region $Q\gg m_c.$ In
practice, this approximation makes little difference when charm production
only contributes a small fraction of the measured structure functions used
in the global analysis. This is no longer the case.

A consistent formulation of PQCD with massive quarks, representing a natural
implementation of the physical principles behind the general formula, Eq.~
\ref{factorlN}, has been given by Aivazis \etal\cite{acot}. It reduces to
the appropriate limits -- FFN scheme near the threshold region ($Q\simeq m_c$%
), and the simple massless QCD formalism in the high energy limit ($Q\gg m_c$%
) -- and provides a unified description of charm production over the full
energy range. This formalism has been applied to the analysis of charged
current lepto-production of charm\cite{ccfr}. Some recent papers have
attempted to compare this approach, referred to as the
variable-flavor-number (VFN) scheme---because the active number of flavors
depends on the scale, with the FFN scheme\cite{ffnvfn}.
Results are not conclusive, since comparable parton distributions\ in the
various schemes are not available in the literature for a consistent
comparison.\footnote{%
In addition, the proper implementation of the VFN scheme requires a delicate
cancellation between three inter-related terms. (See next section.) When the
parton distributions near the threshold
region are not precisely generated to match the scheme,
the cancellation will fail. This hampers most of these references.}

In this paper, we apply the more complete variable-flavor-number formalism
to the global analysis of parton distributions. To carry out a systematic
comparison, we also perform equivalent new analyses in the FFN scheme using
both $n_f=3$ and $n_f=4.$ We compare these results with each other, and with
those obtained previously in the zero-mass approximation (CTEQ4) \cite{cteq}%
. We find that: (i) the more complete formalism gives the best fit to the
global data, further confirming the robustness of the PQCD theory; (ii) the $%
n_f=3$ FFN scheme has difficulty accommodating the hadron collider data
included in the global analysis, whereas the $n_f=4$ fit appears to be
acceptable; and (iii) the recent precision data from HERA at small-$x$ are
also sensitive to the small differences among the various schemes and
approximations. In the concluding section, we discuss the implications of
the improved treatment of heavy quark mass effect in practical applications
of PQCD and compare our results with recent related works.

\section{Treatment of Heavy Quark Mass Effect and New Global Analysis}

We will extend the global QCD analysis to incorporate heavy quark mass
effects according to the formalism of Aivazis \etal \cite{acot}. In
comparison to recent analyses using the massless approach, significant
differences only arise in the charm contribution to deep inelastic structure
functions at small-$x.$ Thus, we will focus on this aspect of the
calculation in most of the following discussions. 

To be consistent with the treatment of all the processes used in the NLO
global analysis, the charm production contribution is calculated to order $%
\alpha _s$---the same as the light-flavor contributions. Thus, the partonic
processes included in the calculation are, for charged current interactions: 
\begin{equation}
\begin{array}{lccccl}
\alpha _s^0: & W^{+}(W^{-}) & + & s(\bar{s}) & \rightarrow  & c(\bar{c}) \\ 
\alpha _s^1:\quad \, & W^{+}(W^{-}) & + & g & \rightarrow  & \bar{s}(s)+c(%
\bar{c})
\end{array}
\label{ccpp}
\end{equation}
and, for neutral current interactions:\footnote{%
Not shown explicitly is the order $\alpha _s$ quark scattering process, \eg $%
\gamma ^{*}(Z)+c\rightarrow g+c$ for NC. Numerically, this process gives a
contribution which is at least an order of $\alpha _s$ down from the ones
shown because $c/g$ $\propto \alpha _s$.} 
\begin{equation}
\begin{array}{lccccl}
\alpha _s^0:\quad \, & \gamma ^{*}(Z) & + & c(\bar{c}) & \rightarrow  & c(%
\bar{c}) \\ 
\alpha _s^1:\quad \, & \gamma ^{*}(Z) & + & g & \rightarrow  & c+\bar{c}
\end{array}
\label{ncpp}
\end{equation}
As pointed out in Ref.~\cite{acot}, the {\em flavor excitation} $\alpha _s^0
$ term and the {\em flavor creation} (or {\em gluon fusion}) $\alpha _s^1$
term cannot be simply added because there is an overlapping region where
they represent the same physics. The formalism provides a consistent
procedure to subtract out the contribution of the collinear configuration in
the gluon fusion process which is already included in the DGLAP evolved
quark distribution in the flavor excitation process. See Fig.~\ref
{fig:diagram} for an illustration. \figdiagram
This procedure eliminates the double counting.\footnote{%
This subtraction has its counter-part in the subtraction of the $\frac
1\epsilon $ pole in the dimensional regularization of collinear
singularities in massless PQCD.} By keeping the quark mass $m_c$ in the hard
cross-sections for all terms, this formalism reduces to the gluon fusion
results of the FFN scheme near threshold because the flavor excitation and
subtraction terms cancel each other in that region. This represents the
correct physics for that region. In the other limit, when $Q\gg m_c$, the
subtraction term cancels the mass singularity of the gluon fusion term, the
theory is free from large logarithms of the FFN scheme, and the perturbation
series provides a good approximation to the correct physics in that region
as well. For details, see Ref.~\cite{acot}. For our purposes, the charm
quark distribution $f^c(x,Q)$ is assumed to be zero at the threshold $%
Q=m_c=1.6$ GeV; and it is dynamically generated by QCD evolution to higher
scales. This approach does allow the option of including the presence of
intrinsic charm inside the nucleon at threshold, which is excluded in the
FFN scheme by definition. This option is of particular interest in the
dedicated study of $F_2^c(x,Q)$\cite{H1c}; it will be explored elsewhere.

Before performing a new global fit, it is instructive to quantify the
difference between the above treatment of the partonic processes for charm
production and the approximation made in previous CTEQ global analyses.
Thus, we first convolute the existing CTEQ4M distributions with the improved
hard-scattering cross-section described above to calculate the structure
functions, and compare the results with the experimental data which were
used in the CTEQ4 analysis. As is known from previous studies\footnote{%
CTEQ notes for the CTEQ3 analysis, 1994 (unpublished).}, very little
difference is found for all the fixed-target experiments. However, the
differences at small-$x$ for the HERA experiments are comparable to the
current experimental errors; hence they do matter. This is illustrated for
three $Q^2$ data bins in Fig.~\ref{fig:hera}. \fighera
The contrast originates from the fact that charm production comprises a few
percent of the measured $F_2$ in the fixed-target energy range; but it rises
to about 25\% at small-$x$ for HERA. This is shown in Fig.~\ref
{fig:percent}. \figpercent
At the few percent level, a relatively large theoretical uncertainty on $%
F_2^c$ can be tolerated, since experimental errors themselves are in the few
percent range. This is no longer the case when the fractional contribution
rises to 25\% with experimental errors in the few percent range. Of
course, it will be even more important to have the theory formulated
accurately in the study of measured $F_2^c$ itself.

These results clearly imply the need to perform new global analyses to
account for the correct physics behind the recent measurements. Thus, we
repeat the CTEQ4 global analysis \cite{cteq}, using the improved theory for
heavy quark production. Charged current and neutral current DIS processes
are treated consistently as described above. We find the overall $\chi ^2$
for the global fit is improved from the previous best fit CTEQ4M---1293 vs.
1320 for 1297 data points, as shown in Table \ref{tbl:chi} where both the
overall $\chi ^2$ and its distribution among the DIS and D-Y data sets are
presented.\footnote{%
Direct photon and jet-production data were also used to constrain the fits.
Because of the difficulty in quantifying current theoretical and
experimental uncertainties for these processes, the specific $\chi ^2$
values are difficult to interpret; hence they are not explicitly presented.
(See Ref.~\cite{cteq} for discussions.)} The small improvement in $\chi ^2$
is spread over both the fixed-target and HERA DIS data sets. The comparison
of this new fit to the small-$x$ data of H1 is included in Fig.~\ref
{fig:hera}. We shall refer to this new set of parton distributions as
CTEQ4HQ (HQ for heavy quark).
\footnote{Computer code for this and the following parton sets will be
available at the CTEQ Web site http://www.phys.psu.edu:80/\~{}cteq/.}

As expected, the deviation of CTEQ4HQ distributions from CTEQ4M are rather
minor, and they are most noticeable at small-$x.$ Fig.~\ref{fig:pdf} 
\figpdf
shows the comparison of these parton distributions. Interestingly, the
differences are more obvious for the light quarks than for the gluon and
charm; and they consist of an increase in these distributions throughout the
small-$x$ range (which is most visible in the plot). This is because the
improved theory for heavy quark production reduces the predicted
cross-section which has to be brought back to the experimental values by
increased parton distributions. This can be done most efficiently by a small
fractional increase in the light quark flavors. Of course, if the momentum
sum rule is to be preserved, there has to be a slight decrease in the parton
distributions at large $x.$ This is indeed the case, even though it is not
quite visible in this plot. Proportionally, this decrease can be very small,
since the momentum sum-rule integral strongly suppresses differences at
small-$x.$\tblchi

\section{Global Analyses in the Fixed-Flavor-Number Scheme}

Since much of the current literature on heavy flavor production in recent
years adopts the fixed-flavor-number scheme described in the introduction%
\cite{GHR,GRV,smithetal,ffnvfn}, 
it is useful to have available up-to-date parton distributions in this scheme 
and to study its efficacy in describing
existing global data. By comparing results from such an analysis with those
obtained above, one can also draw more definitive conclusions concerning the
effectiveness of the FFN scheme versus the more complete
variable-flavor-number formalism.

The FFN scheme is simpler to implement, since it includes only light partons 
$\{g,u,d,s\}$ in the initial state. Thus, for neutral current deep inelastic
scattering as an example, the only partonic process contributing (to order $%
\alpha _s\,)$ is the gluon fusion mechanism $\gamma ^{*}(Z)+g\rightarrow c+%
\bar{c}.$ There is no need for any subtraction, since the charm mass
regulates the collinear singularity, and there is no double counting. Using
this scheme consistently for all processes, one can perform a global
analysis in the $n_f=3$ FFN scheme, and obtain an appropriate set of parton
distributions. As in the previous section, the treatment of data sets and
the global analysis procedures are the same as in the previous CTEQ4
analysis \cite{cteq}. We shall call this the CTEQ4F3 set.

Although we have emphasized charm production due to its immediate relevance,
the same physics issues apply to bottom quark production. At HERA, $b$
production is quite negligible in the total $F_2.$ But at LEP, the
Tevatron and LHC, it is not. For this reason, and due to the fact that $m_b$ is
genuinely ``heavy'' whereas $m_c$ is really on the borderline, we also carry
out a global analysis in the $n_f=4$ FFN scheme for completeness. For this
case, charm is counted as an active parton and treated above threshold in
the manner described in the previous section, but bottom is treated as a
heavy quark in the spirit of the FFN scheme. The set of parton distributions
obtained this way is called CTEQ4F4.

An overview of the comparison between these FFN scheme global fits to the
ones described in the previous section is included in Table\ \ref{tbl:chi}.
We see that the overall $\chi ^2$ for the two FFN scheme fits are 1380 and
1349 (for $n_f=3,4$ respectively) compared to 1293 for CTEQ4HQ. A
substantial part of the difference is due to the HERA data (particularly the
ZEUS data set). Other than that, the $n_f=4$ FFN fit (CTEQ4F4) is very close
to both CTEQ4M and CTEQ4HQ. This is expected since the only difference lies
in the treatment of the small bottom quark contribution to the measured
quantities.

More revealing is the $n_f=3$ FFN fit (CTEQ4F3) which shows some signs of
difficulty in achieving a good fit. In addition to the increased $\chi ^2$
for the HERA data, there are two possible problems with hadron collider
data: (i) the $\chi ^2$ on the CDF $W$-lepton asymmetry data \cite{Wasym}\
increases slightly; and, not shown in this table but more significantly,
(ii) this fit requires both the CDF and D0 inclusive jet production data
sets \cite{jets} to be normalized down by 10\% for a reasonable fit. 
Because of the absence of $c$ and $b$
partons inside the nucleon in the $n_f=3$ FFN scheme (by definition), the
light quark distributions have to be substantially increased in order to
compensate. This distorts the $u$ and $d$ distributions with respect to the
standard parton sets. The $d/u$ ratio is known to be important in
determining the $W$-lepton asymmetry. The latter is also affected by the
different treatment of charm parton. \cite{GRV}
Since both points (i) and (ii) concern
high energies, they suggest the neglect of important physics at large energy
scales in this scheme. At the $M_W	$ scale, it surely is not a good idea to
treat the charm quark as a ``heavy quark''. It may be argued that the
inclusion of higher order ($\alpha _s^2\,)$ terms could improve the
agreement between theory and experiment. (We have not done this in order to
keep the discussions clear, and to keep all calculations to the same order.)
This may be true. But that will only postpone the same problems to higher
energy ranges. In fact, as mentioned above, the $M_W	$ scale is already here
with us; any temporary gain in using a scheme appropriate for the threshold
region has to be weighed in that broader context.

With these provisoes, the CTEQ4F3 and CTEQ4F4 do represent the most up to
date parton distributions to be used with FFN scheme calculations. CTEQ4F3
is in the same scheme as GRV94\cite{GRV}. CTEQ4F3 gives better fits to the
global data, and are comparable to the other CTEQ4 distributions for the
purpose of studying scheme dependences. These distributions can be further
improved with the inclusion of order $\alpha _s^2$ hard cross-sections and
with more recent data.\footnote{%
Since the publication of the previous CTEQ4 analysis, NMC has published
results of their final data analysis \cite{nmc96}. The differences between
the old and new results are not significant. We did not replace the previous
NMC data by the new ones in this analysis because, in order to demonstrate
the small differences among the various schemes, including the previous
CTEQ4M analysis, we have to use the same data sets. We also anticipate new
data on the CCFR measurement of the charge current structure functions
in the near future.}

\section{Concluding Remarks}

Recent HERA data both on precision measurement of the inclusive $F_2(x,Q)$
and on $F_2^c(x,Q)$ require a more careful theoretical treatment of heavy
quark production in PQCD. This entails, in turn, both a clearer definition
of the perturbative scheme used and the careful choice of a scheme which is
appropriate for the full energy range probed by the experiments in question.
We have described the salient features of most of the schemes which have
been used (explicitly or implicitly) in current literature, and made a
systematic comparison in the context of the global QCD analysis of hard
processes to extract the universal parton distributions.

The advantages of both the massless approximation to the
variable-flavor-number scheme used by most previous global analyses (\eg
both CTEQ and MRS ) and the FFN scheme used by previous heavy quark
production calculations (as well as by the GRV parton sets) lie in their
simplicity in implementation. They both have limitations which can no longer
be totally neglected. The generalized variable-flavor-number formalism
contains the right physics in the full energy range, including both the
threshold and asymptotic regions, and is free of large logarithms (except
those of the small-$x$ kind which we have not discussed). It is reassuring
to see that the more complete theory does give a better description of the
wide range of experimental data included in the global analysis; and,
non-trivially, the less complete theory does fall short where it is expected
to. This provides further confirmation of the soundness of the PQCD theory.

Since the completion of this study, a preprint on a new treatment of
heavy quark production appeared\cite{mrrs}. This (MRRS) approach is also in
the variable-flavor-number scheme and is the same in spirit to the formalism
of Aivazis \etal which we adopt, hence it shares the characteristics
mentioned in the previous paragraph. The new MRRS parton set is comparable
to CTEQ4HQ. The differences in implementation of the basic ideas of the VFN
scheme lie in the specific choice of the location of the charm threshold and
in the detailed treatment of mass effects in the region just above threshold.

We note that, the increase in precision of experiments and in sophistication
of the theory place more demand on users of the QCD parton formalism: {\em %
for consistency, each set of new parton distributions can only be applied to
hard scattering cross-sections calculated in the same scheme}. When charm
and bottom mass effects matter, one must be fully aware of which heavy quark
mass scheme should be used in a given calculation -- this is {\em in
addition to} the familiar $\overline{MS}$ and DIS schemes of the massless
theory. For most lepton-hadron and hadron-hadron processes away from very
small-$x$, data cannot distinguish among the various schemes studied in this
paper, provided the parton distributions and hard scattering calculations
are not mismatched. In these cases, the usual parton distributions, even if
not precisely the true ones of nature, provide an effective description of
the physics probed. In this sense, the continued use of existing standard
parton distribution sets (such as CTEQ \cite{cteq} and MRS\cite{mrs}) for
these processes is acceptable. However, for processes sensitive to initial
or final state heavy quarks, it will be imperative to use the more complete
theory, with matching parton distributions, if meaningful physical
quantities are to be extracted. For these processes, current theory can
still be improved to include higher order terms; and detailed phenomenology
is yet to be done when both experiment and theory mature.

\vspace{1cm}

Acknowledgments: We thank John Collins, Frederick Olness and Carl Schmidt
for useful discussions on heavy quark physics; Joey Huston, Steve Kuhlmann,
Joseph Owens, Davison Soper and Harry Weerts for collaboration on the
previous CTEQ4 analysis on which the current study is based; and Kuhlmann
and Raymond Brock for detailed comments on the manuscript.

\end{document}